\documentclass[conference]{IEEEtran}
\IEEEoverridecommandlockouts

\usepackage{cite}
\usepackage{amsmath,amssymb,amsfonts}
\usepackage{algorithmic}
\usepackage{graphicx}
\usepackage{xcolor}
\def\BibTeX{{\rm B\kern-.05em{\sc i\kern-.025em b}\kern-.08em
T\kern-.1667em\lower.7ex\hbox{E}\kern-.125emX}}

\graphicspath{ {./figs/} }

\title{SecureCyclon: Dependable Peer Sampling}

\usepackage{alphabeta}  

\usepackage{hyperref}
\hypersetup{
    colorlinks=true,
    linkcolor=blue,
    citecolor=magenta,
    filecolor=magenta,
    urlcolor=cyan,
    pdftitle={SecureCyclon}
}
\usepackage{ifthen}
\usepackage{calc}
\usepackage{color}
\usepackage[normalem]{ulem}
\usepackage{tikz}
\usepackage{amsfonts}
\usepackage{pgfplots}

\newcommand{\removed}[1]{}

\DeclareRobustCommand{\change}[4]{{\color{#3}#1}\ifx\relax#4\relax\else\xspace\textcolor{Gray}{\sout{#4}}\fi}

\newcommand{\figref}[1]{Figure~\ref{#1}}
\newcommand{\secref}[1]{Section~\ref{#1}}

\newcommand{\prot}{\textsc{SecureCyclon}}

\author{\IEEEauthorblockN{Alexandros Antonov}
\IEEEauthorblockA{\textit{Department of Informatics} \\
\textit{Athens University of Economics and Business}\\
Athens, Greece \\
aantonov@aueb.gr}
\and
\IEEEauthorblockN{Spyros Voulgaris}
\IEEEauthorblockA{\textit{Department of Informatics} \\
\textit{Athens University of Economics and Business}\\
Athens, Greece \\
voulgaris@aueb.gr}
}

\newcommand\copyrighttext{%
    \footnotesize \textcopyright 2023 IEEE.  Personal use of this material is permitted.  Permission from IEEE must be
    obtained for all other uses, in any current or future media, including reprinting/republishing this material for
    advertising or promotional purposes, creating new collective works, for resale or redistribution to servers or lists,
    or reuse of any copyrighted component of this work in other works.
    DOI: 10.1109/ICDCS57875.2023.00041}
\newcommand\copyrightnotice{%
    \begin{tikzpicture}[remember picture,overlay]
        \node[anchor=south,yshift=10pt] at (current page.south) {\fbox{\parbox{\dimexpr\textwidth-\fboxsep-\fboxrule\relax}{\copyrighttext}}};
    \end{tikzpicture}%
}

\begin{document}

    \maketitle
    \copyrightnotice

    \begin{abstract}

Overlay management is the cornerstone of building robust and dependable Peer-to-Peer systems.
A key component for building such overlays is the peer-sampling service, a mechanism that continuously supplies each node with
a set of up-to-date peers randomly selected across all alive nodes.
Arguably, the most pernicious malicious action against such mechanisms is the provision of arbitrarily created links that point at malicious nodes.
This paper proposes \prot, a peer-sampling protocol that deterministically eliminates the ability of malicious nodes to overrepresent themselves in Peer-to-Peer overlays.
To the best of our knowledge, this is the first protocol to offer this property, as previous works were able to only bound the proportion of excessive links to malicious nodes, without completely eliminating them.
\prot\ redefines the concept of node descriptors from just being containers of information that enable communication with
specific nodes, to being communication certificates that traverse the network and enable nodes to provably discover malicious nodes.
We evaluate our solution with the conduction of extended simulations, and we demonstrate that it provides resilience even at the extreme condition of $40\%$ malicious node participation.

\end{abstract}

\begin{IEEEkeywords}
peer sampling, network overlays, gossip, protocol enforcement
\end{IEEEkeywords}

    \section{Introduction}
\label{sec:introduction}

Gossiping has shown to be a promising paradigm for a variety of Peer-to-Peer (P2P) applications in large-scale decentralized systems, most notably for information dissemination~\cite{LightweightEpidemicDissemination,LightweightProbabilisticBroadcast}~,
data replication~\cite{EpidemicDatabaseMaintenance}, data aggregation, overlay construction~\cite{voulgaris.jnsm.2005,EpidemicRoutingTableManagement,Vicinity}, fault detection~\cite{vanrenesse.middleware.1998}, and decentralized clustering.
This has been emphatically stressed in recent years through the advent of blockchain systems, most of which rely on gossiping for efficiently and reliably disseminating blocks to thousands of nodes spread around the globe~\cite{Gossipsub,perigee,Cougar}.

A fundamental operation underlying gossiping protocols is the ability of nodes to pick gossiping partners at random, an operation known as \emph{peer sampling}.

Picking gossiping partners at random is essential to the operation of most gossiping protocols.
For instance, dissemination protocols need news to spread to all nodes with equal probability.
Aggregation protocols need all participants' data to be equally represented in the computed aggregate, and all nodes to equally contribute to its calculation.
Clustering algorithms need a uniform sampling of nodes in order to match them in optimally formed clusters.
Fault detection algorithms require that nodes be monitored by an unbiased selection of other nodes to properly detect faulty behavior.
Finally, by contacting nodes with uniform probability, load is more evenly shared among them.

Equally importantly, random connections also play a crucial role for the robustness of the overlay network.
A set of nodes form an \emph{overlay network} by establishing logical links with each other.
When these links are selected uniformly at random, the emerging overlays resemble random graphs.
Random graphs are known for their remarkable robustness, in the sense that they remain connected in a single component even when the majority of nodes are removed.
To this end, creating an overlay network with links between nodes set at random helps nodes stay connected in a robust, single-component overlay that remains connected even in the face of high node churn or catastrophic failures \cite{voulgaris.jnsm.2005}.

As simple as it may sound, random peer sampling is in fact far from trivial, in particular when massive-scale, dynamic, decentralized scenarios are at stake, where nodes are not in a position to maintain a complete and up-to-date view of the entire network.

Early efforts to discover random nodes in a network resorted to methods based on random walks.
Such methods have not only been employed in early P2P systems, such as Gnutella~\cite{Gnutella}, but also in early blockchains, most notably in Bitcoin~\cite{Bitcoin}.
Random walks, however, suffer from a number of issues.
First, they have shown to favor the discovery of high-degree nodes, resulting in overlays with highly imbalanced node degrees (hubs vs poorly connected nodes), if used to discover neighbors for newly joined nodes.
Second, sampling peers through random walks is an inherently reactive procedure, failing to follow the continuously changing dynamic membership in a real-world network.
Third, they could easily be exploited by malicious nodes.
For instance, a relatively small number of instrumented Bitcoin clients could collude to keep random walks that reach any one of them confined within their community, gradually directing most of Bitcoin traffic through them.
This would put them in a position to filter out transactions or blocks that are against their interests, or to cause wreak havoc by blocking traffic altogether.

An explicit family of peer-sampling protocols has emerged to tackle these specific issues, namely, random peer sampling and robust overlay maintenance.
Protocols of the peer-sampling family follow a simple operational pattern, outlined below.
Each node maintains links to a small and fixed number (e.g., 20 to 50) of neighbors, referred to as the node's \emph{view} of the network.
Each node periodically initiates a push-pull exchange with one of its neighbors, and the two nodes send each other their current views or parts of them.
Upon receiving the counterparty's view, a node updates its own by combining the newly received neighbors with the ones it already had, keeping no more than the fixed number of neighbors its view is configured to hold.
This way, each node's view is periodically being refreshed, effectively providing the node with a continuous stream of random neighbors from the entire set of alive nodes.

Different policies on how to mix a received view with the local one, which neighbor to gossip with, which neighbors to trade, etc., lead to different instances of the peer-sampling family, differing in how fast peers join the overlay, how fast stale links to departed peers are removed, how efficiently the overlay heals from failures, and more \cite{PeerSampling}.
By and large, however, all peer-sampling protocols provide each node with a continuous stream of uniformly randomly selected samples of all alive peers, maintain overlays with remarkable robustness and self-healing properties, and are extremely scalable, thanks to the seamless cooperation of all participating nodes.

Unfortunately, it is exactly this reliance of peer-sampling protocols on the cooperation of participating nodes that renders them highly vulnerable to malicious behavior.
Peer-sampling protocols can become easy targets for malicious participants that deviate from the protocols' prescribed operation to promote biased outcomes.
Similarly to the case of exploiting random walks outlined earlier, a small group of malicious peers could easily take over all links of the network, namely by consistently presenting to legitimate peers views with links exclusively to members of their group.
Legitimate peers' views will gradually get contaminated by increasingly more links to malicious peers at the expense of links to legitimate ones, which would further facilitate and accelerate the attacker's job.

In no time, all legitimate peers would point exclusively at malicious ones.
This would give the attacker complete control over all traffic flowing through the network and on legitimate nodes' connectivity.
In the extreme form of this attack, malicious peers could abandon the system in a coordinated move, letting the network segregated into a large number of single-node disconnected components.
This has been coined as the \emph{hub attack} in~\cite{SecurePeerSamplingService}.

In this paper, we are addressing the hub attack focusing on Cyclon~\cite{voulgaris.jnsm.2005}, a popular peer-sampling protocol.
We propose a solution that not only detects and negates the effects of malicious behavior in a timely fashion, but also provides indisputable proof of protocol misconduct, allowing the network to blacklist and remove cheating nodes.


The rest of the paper is organized as follows.
\secref{sec:background} provides the necessary background, describing our system model, the legacy Cyclon protocol, and the attack model.
\secref{sec:challenges} lays out the challenges we are addressing.
\secref{sec:protocol} presents and advocates our protocol's design, explaining its operation and supporting the design decisions taken.
In \secref{sec:protocol-integration} we tackle the problems that arise from merging our protocol with Cyclon.
\secref{sec:evaluation} evaluates our protocol.
Finally, \secref{sec:related-work} surveys related work, and \secref{sec:conclusions} concludes.

    \section{Background}
\label{sec:background}

In this section, we provide the necessary groundwork that is needed to follow the rest of this paper.
Initially we describe the system model, which introduces the basic notions, define the terminology, and outline the assumptions used in our work.
Subsequently, we delve into Cyclon, the peer-sampling protocol that is the focus of our research.
Our Attack model concludes this section, which defines the rationale behind the actions of malicious nodes.

\subsection{System Model}
\label{subsec:system-model}

We consider a network of $n$ nodes, each having a unique ID.
Nodes are connected over a routed network infrastructure, which allows communication between any pair of them.
The sole condition for communication is that the sender knows the network address (e.g., the IP address and port) of the receiver.

Information about neighbors is stored and exchanged by means of \emph{node descriptors}, also referred to as \emph{links}.
A node descriptor contains the node's unique ID and its network address.
The descriptor of a node may be generated exclusively by the node itself, but it can be freely passed around from any node to any other.

The network is dynamic, nodes may join, leave, or fail, with no prior notice.
Messages may be delayed or dropped.
However, message integrity is guaranteed and malicious nodes cannot impersonate legitimate ones, as every message is cryptographically signed by its sender.

Network nodes possess clocks that are relatively synchronized with each other.
We use \emph{cycles} as a time unit in protocol design and evaluation.
A cycle corresponds to the period during which a node is allowed to initialize exactly one gossip exchange.

Each node has exactly one cryptographic private/public key pair.
We set the value of the unique ID of each node to be equal to the value of its public key.

We assume that the acquisition of unique identifiers is not a trivial process.
This property can be achieved by mechanisms described in~\cite{TheSybilAttack}, such as relying on a trusted authority, or having to solve a unique computational puzzle in order to acquire an identifier.

\subsection{Cyclon}
\label{subsec:cyclon}

Cyclon~\cite{voulgaris.jnsm.2005} is a popular peer-sampling protocol, known for building overlays that share a lot of properties with random graphs.
Cyclon is remarkably scalable, robust to failures, and lightweight.
It operates in a fully decentralized, self-organizing fashion.

In Cyclon, node descriptors contain the following fields:
\begin{itemize}
  \item The node's ID
  \item The node's network address
  \item A timestamp denoting when this specific descriptor was generated
\end{itemize}

Each node maintains a small partial view of the network of size $\ell$ (known as the \emph{view length}), that is, a list of $\ell$ descriptors of other nodes.
Periodically, a node selects the \emph{oldest} descriptor in its view, removes it from its view, and initiates a gossip exchange with the respective peer.
In this gossip exchange, the initiator first sends $s$ descriptors to its gossip partner: a fresh descriptor of itself plus another $s-1$ descriptors selected (and removed) from its view at random.
Upon receiving these, the gossip partner responds by also sending $s$ descriptors back, all of them selected (and removed) from its view at random.
Each node stores the $s$ received descriptors in its view, filling in the gaps from the $s$ descriptors it removed.
In case a node finds itself with empty slots in its view after an exchange (e.g., because it has recently joined and its view is not fully populated yet, or because the other party did not provide enough descriptors), it is free to retain the descriptors it sent to the other party.

In a Cyclon gossip exchange, nodes effectively \emph{swap} some of their descriptors, hence, parameter $s$ is known as the \emph{swap length}.
By doing so, overlay links are being mixed at random, resulting into overlays that demonstrate remarkable similarity to random graphs, as has been shown in~\cite{voulgaris.jnsm.2005}.

\begin{figure}
  \includegraphics[width=\linewidth]{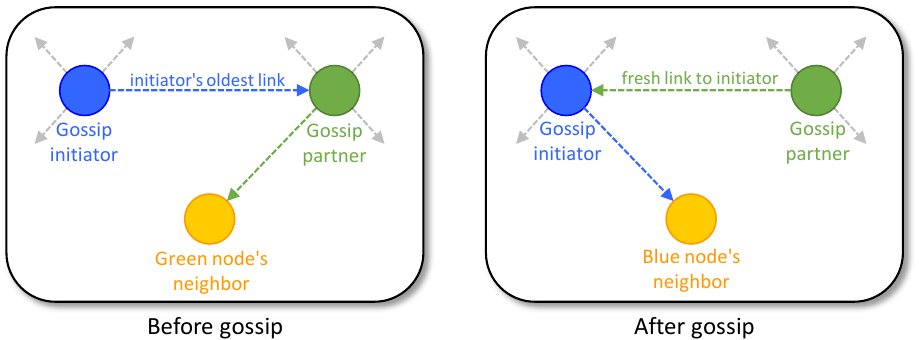}
  \caption{Gossip exchange in Cyclon. The initiator (blue) redeems its link to its gossip partner (green), replacing it by a link to a third node (yellow) provided by the partner. The gossip partner removes its link to a third node, replacing it by a fresh link to the initiator provided by the initiator. Note: The two nodes may also exchange a few more random links, one-to-one, not shown here.}
  \label{fig:cyclon_gossip}
\end{figure}

Note that when a node initiates gossiping to a neighbor, it removes that neighbor's descriptor from its view.
We say that it \emph{redeems} that descriptor for a gossip exchange.
Given a fail-free environment, the overlay connectivity is nevertheless preserved, as the neighboring relation between the two nodes is not lost, but simply changes direction.
Links to third nodes are not removed either.
Third nodes just move from being neighbors of one node to being neighbors of another.
\figref{fig:cyclon_gossip} illustrates a gossip exchange between two nodes.

Cyclon renews the links of the overlay by removing the oldest descriptors and by injecting a fresh descriptor per node per cycle.
This way, a descriptor's life expectancy becomes naturally bounded, as a descriptor is not likely to get redeemed before it becomes old, while it is also not likely to become too old before it gets redeemed.
In combination with the fixed birth rate of descriptors, this implies that any given node's descriptors will have a relatively stable population over time.

Interestingly, the descriptor population of each individual node is inherently being pushed towards an equilibrium.
A node \emph{increases} its indegree by one (by injecting a fresh descriptor) precisely once per cycle, when it initiates a gossip exchange.
On the other hand, a node's indegree is \emph{decreased} by one when it is chosen as a partner in a gossip exchange initiated by someone else (someone who redeems one of the node's old descriptors).
Therefore, a node with a low indegree (i.e., few descriptors floating around) is likely to be contacted less than once per cycle, on average, therefore its descriptors will be getting redeemed slower than they are getting born, thus the node's indegree will tend to rise.
On the opposite side, a node with a high indegree (i.e., lots of its descriptors around), will be contacted more than once per cycle, on average, seeing its descriptors being redeemed at a faster pace than being born, thus the node's indegree will tend to drop.

\begin{figure}
  \includegraphics{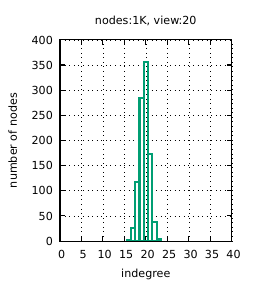}
  \includegraphics{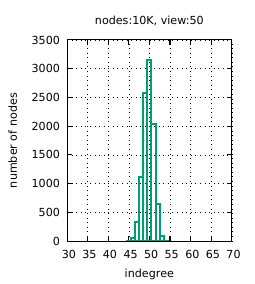}
  \caption{Indegrees in Cyclon overlays. Each individual node's indegree is very closely bounded to the system-wide configured outdegree.}
  \label{fig:degree}
\end{figure}

The consequence of the aforementioned observations is that, besides having a fixed outdegree as explicitly configured through their view length, nodes in Cyclon also experience a tightly bounded indegree, which is shown to inherently fluctuate around its outdegree (i.e., its view length $\ell$), with very small deviation.
\figref{fig:degree} presents the indegree distribution of two Cyclon overlays, for network sizes of 1K and 10K nodes, respectively.
These overlays show that they are extremely robust, as no node is left behind with a low indegree.

\begin{figure*}[t]
  \includegraphics{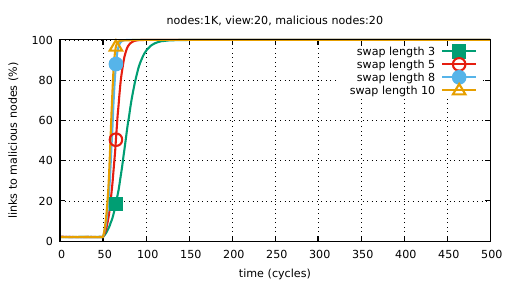}
  \includegraphics{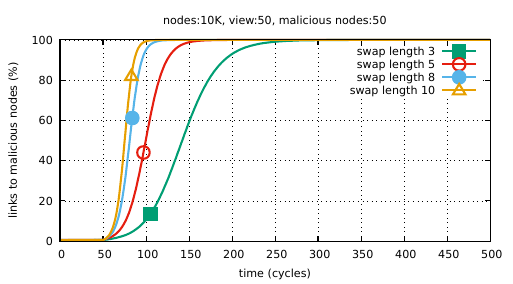}
  \caption{A small set of malicious nodes can easily take over 100\% of the network.
  These experiments of 1K nodes with view length 20 (left) and 10K nodes with view length 50 (right), demonstrate how fast an attacker can take full control of the network by deploying even the minimum number of malicious nodes required, namely, 20 and 50, respectively.
  In these experiments, all nodes operate correctly until cycle 50, during which the percentage of links to malicious nodes is analogous to their population, as expected.
  From that point onwards, malicious nodes continue gossiping seemingly correctly (i.e., correct rate, correct gossip exchanges, etc.), however they present a view consisting of malicious  nodes exclusively.}
  \label{fig:hubattack}
\end{figure*}

Despite its desirable properties, Cyclon comes with a major shortcoming, namely its vulnerability to malicious nodes.
Cyclon has been designed around the assumption of seamless cooperation between nodes.
When this is not the case, it becomes an easy target to attacks.
A handful of colliding malicious nodes that deviate from the protocol's rules, may easily leverage Cyclon's gossip exchanges to take over all links to and from legitimate nodes.
It is not hard to imagine that by presenting to legitimate nodes fake views containing only links to their malicious colleagues, malicious nodes can gradually pollute legitimate nodes' views, eventually taking over 100\% of the links.

\figref{fig:hubattack} demonstrates this vulnerability.
It presents experiments for a 1K and a 10K-node network, in which a small number of nodes, equal to the view length $\ell$, manage to take over all links in a very short number of cycles.
In all experiments, malicious nodes behave correctly until cycle 50, after which they start their coordinated attack.

Our work proposes \prot, a revised, security-sensitive extension of Cyclon, which efficiently shields Cyclon overlays from such attacks.

\subsection{Attack Model}
\label{subsec:attack-model}

In our attack model, we assume a party of malicious nodes that run under the same operator.
Malicious nodes collude with each other, have mutual knowledge about the network, share the same pool of node descriptors, and forge node descriptors on demand to assist each other.


The ultimate goal of the malicious party is to monopolize all the connections from and to legitimate nodes.
In pursuit of this objective, each individual malicious node is required to pollute the network with descriptors that point to nodes of its malicious party, ultimately establishing hub positions as described in \ref{subsec:the-hub-attack}.

As pointed out in~\ref{subsec:cyclon}, when Cyclon is employed, the links of the overlay are continuously recycled, wherein old links are replaced by newly generated ones.
Taking this into consideration, a malicious party would not reach its goal if its participants are slow to pollute the network.
In order to outperform the self-healing properties of Cyclon, malicious nodes employ aggressive strategies that involve rapid provision of supplementary node descriptors.

As showcased in~\cite{voulgaris.jnsm.2005}, Cyclon demonstrates remarkable efficacy in redistributing node descriptors throughout the network, thereby equipping each overlay node with a randomized set of peers.
Even if a malicious node communicating with a legitimate node memorizes the view of its victim, the view will change in a span of a few cycles.
Due to this, malicious nodes in our model do not endeavor to predict the contents of legitimate nodes' views, when choosing which of their peers they will communicate with, or when selecting which nodes from their party the supplementary descriptors will point to.
Instead, they make such choices uniformly at random.

    \section{Challenges}
\label{sec:challenges}

Despite the desirable characteristics exhibited by gossip-based peer-sampling protocols, managing a dynamic set of neighbors proves to be an easy target to adversary behavior.
More specifically, an attacker could apply one or more of the following protocol violations to take over a Cyclon overlay:
\newline

\noindent
\textbf{Frequency Violations:} By initiating gossip exchanges faster than the prescribed period, a node could generate descriptors to itself at a higher frequency than other nodes, effectively increasing its overall link representation in the network.

Gossiping slower, on the other hand, does not pose any threat to the network, and should not be considered an attack.
After all, it may be attributed to natural causes, such as network delays or node disconnections.

\noindent
\textbf{Partner Selection Violations:} Instead of selecting a gossiping partner as prescribed by the protocol, a node may arbitrarily select another node that better serves the attacker's strategy.

\noindent
\textbf{View Violations:} The most impactful protocol violation derives from a node's freedom to present any arbitrary set of descriptors as its view during a gossip exchange, and to arbitrarily select which descriptors to send to the other party.
It may send fewer descriptors than supposed to, it may send descriptors that are not present in its current view, or it could even create fake node descriptors on the spot.

For example, malicious nodes could selectively present descriptors of malicious nodes to their legitimate peers, rather than randomly chosen ones from their views, leading to the hub attack demonstrated in the experiments of \figref{fig:hubattack}.
\newline

The aforementioned protocol violations can be used as building blocks to orchestrate high-impact, large-scale attacks, which may disturb the canonical operation of individual nodes or even harm the integrity of the overall network.

Two high-scale attacks that have been extensively studied in the literature are the Hub Attack~\cite{SecurePeerSampling} and the Eclipse Attack~\cite{TheEclipseAttack}.

\subsection{The Hub Attack}
\label{subsec:the-hub-attack}

In this attack, the attacker pollutes the network with node descriptors that point at malicious nodes.
The attacker's goal is to redirect as many links of legitimate nodes as possible to malicious nodes controlled by him.
In a successful (for the attacker) hub attack, malicious nodes become hubs in the overlay.
After establishing central positions, the malicious nodes can leave the overlay, splitting the overlay graph into many disjoint components.
This attack results in a massive DoS attack that can only be reversed by reconstructing the entire network overlay from scratch.

\subsection{The Eclipse Attack}
\label{subsec:the-eclipse-attack}

In this attack a malicious party targets specific nodes, trying to take control of all their overlay connections.
By isolating the targeted node(s) from the rest of the network, the malicious party gains control over all communications of the victim(s),
being able to delay or drop messages from and to them.

Brahms~\cite{Brahms} proposes a generic solution against eclipse attacks.
Each node applies a secret ordering on the stream of node IDs it receives through gossiping, in order to render a small part of the view immune to overprovisioning of malicious node descriptors.
It is considered that by maintaining links to a uniformly random selection of all IDs seen so far, at least some of these links should correspond to legitimate nodes, with high probability.
However, this is not in line with peer-sampling protocols' goal of maintaining views up to date with fresh links to alive peers.

A more centralized approach would be to establish some connections to nodes that are widely trusted.
For example, in the Cardano blockchain~\cite{cardano}, where Stake Pool Operators (SPOs) are considered trusted, and whose nodes are publicly known and registered on the blockchain itself, each network node could dedicate some of its connections to SPO nodes.

Finally, in certain cases it may be possible for nodes to locally detect whether they have been eclipsed and to take corrective steps.
For example, in the Cardano blockchain again, nodes can locally detect whether they have been eclipsed by observing the frequency and contents of received blocks.
When a node determines it is in eclipsed state, it simply leaves the network and joins anew.
Such a mechanism is described in~\cite{CardanoShelley}.
However, such solutions focus on the \emph{detection} of attacks rather than on their \emph{prevention}, and are inherently application-specific, lacking generic applicability.

\subsection{Hub Attack vs Eclipse Attack}
\label{subsec:hub-attack-vs-eclipse-attack}

It is important to note the orthogonality between these two attacks.
An approach that addresses eclipse attacks does not constitute a protection against hub attacks, and vice versa.

In approaches that rely on eclipse detection, malicious nodes could simply run in stealth mode, exhibiting legitimate behavior concerning their application-specific actions.
This strategic approach persists until they have achieved the required link over-representation, at which moment they can deploy a hub attack on the network.
But even when approaches that attempt to prevent eclipse attacks are in place, an attacker having acquired an overwhelmingly high fraction of the links could still deploy a hub attack and break the overlay into many small disjoint components, even if no single node has individually been eclipsed and fully isolated from other legitimate nodes.

On the other hand, a network encompassing mechanisms shielding it from hub attacks, cannot automatically guarantee that no single node will be targeted and will become individually eclipsed, assuming that malicious nodes have the possibility to deviate from the protocol's rules.

    \section{The \prot\ Protocol}
\label{sec:protocol}

The novel idea of \prot\ is that it redefines the concept of \emph{node descriptors}, from simple data structures carrying node metadata, to  unique, unforgeable, and unclonable tokens required to participate in gossiping.
Through this novel concept, adversaries that attempt to increase their share of overlay participation by either generating descriptors faster than allowed, or by cloning existing descriptors, can be discovered by correct nodes.
Most importantly, correct nodes are able to provide indisputable proof of the violations they discover, enabling the entire network to hold violators accountable of their actions, and to blacklist them from all future communication.

That said, \prot\ remains a protocol of predominantly probabilistic nature, in the sense that not every single violation is guaranteed to be discovered, however, malicious nodes that systematically violate the protocol \emph{will} eventually be caught \emph{and} blacklisted.
This is fully in line with the original Cyclon protocol's self-healing properties, as sporadic violations do not constitute a threat to the health of the overlay.
Instead, malicious nodes should rapidly and repetitively violate the rules in order to deploy the high-impact attacks discussed in \secref{sec:challenges}, in an attempt to outperform Cyclon's self-healing properties.
As we demonstrate in our evaluation, when paired with proper configuration, \prot\ can withstand coordinated attempts to conduct malicious actions.

It should be mentioned that \prot\ relies on the collective effort of all correct nodes participating in the network.
We assume that rational nodes value the importance of security, hence will not free-ride on the effort of others, especially given that the protocol has very reasonable resource demands.
On the other hand, it is assumed that malicious nodes opt not to participate in the protocol's security enhancements, as this would work against their own interest.

\subsection{Enhanced node descriptors}
\label{subsec:enhanced-node-descriptors}

\begin{figure}
    \includegraphics[width=\linewidth]{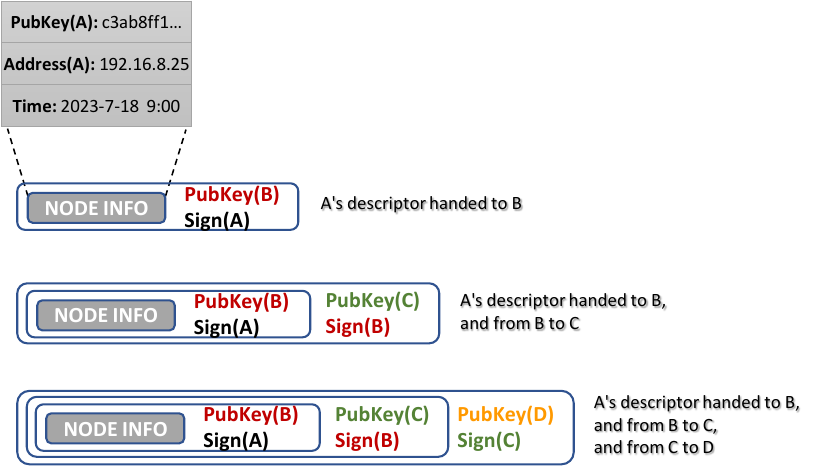}
    \caption{Node $A$ generates a fresh descriptor of itself, and transfers it along with its ownership to node $B$.
    Subsequently $B$ transfers the descriptor and its ownership to node $C$,
    which, in turn, transfers it to node $D$.
    The node descriptor carries the chain of ownership ($A\rightarrow B\rightarrow C\rightarrow D$) of its entire lifetime.}    \label{fig:descriptor}
\end{figure}

A Cyclon node descriptor goes through three phases in its lifespan, its \emph{creation}, its \emph{traversing}, and its \emph{redemption}.
A descriptor's \emph{creation} is when it gets initially established by its \emph{creator}, the node it points at.
The \emph{traversing phase} starts when its creator transfers it for the first time to another node during a gossip exchange.
The descriptor may be transferred further on during this phase, in the context of subsequent gossip exchanges.
The descriptor's \emph{redemption phase} is when the node currently holding that descriptor selects it to initiate a gossip exchange to the node the descriptor links to, that is, the descriptor's creator.
The redemption corresponds to the action of presenting the descriptor back to its creator in order to be allowed to gossip with that specific node.
At the end of this phase, the lifecycle of the node descriptor ends, and it gets permanently removed.

In our enhanced version of node descriptors, a descriptor holds the following information.
When created, a descriptor stores its creator's public key (which also serves as its ID), its network address, and a wall-clock timestamp.
Our protocol does not enforce strict clock synchronization between nodes, other than keeping their clocks updated in a best-effort fashion, e.g., through an ordinary NTP service.
Nevertheless, each node should review the newly created descriptors it receives and reject the ones that contain a timestamp with a high deviation from its own clock.
A descriptor's timestamp is used to discover frequency violations, as we will see in the following section.

During the traversing phase, when a descriptor is passed on from one node to another, we say that the receiving node acquires the descriptor's \emph{ownership}.
The initial owner of a node descriptor is its creator.
A node $A$ transfers a descriptor's ownership to a node $B$ by appending $B$'s public key to the descriptor, and by then signing the entire data structure with its own private key.
By continuously appending owners to a descriptor, a chain-like structure is formed that can track the descriptor's predecessors all the way up to its creator.
We call this structure the \emph{chain of ownership}, and it is used to discover cloned descriptors, as we will see in the following section.
\figref{fig:descriptor} illustrates a descriptor and three consecutive transfers of its ownership.

The final step of our protocol takes place in the redemption phase.
In order for a node to be allowed to initiate a gossip exchange with one of its neighbors, it must send to that neighbor a descriptor for which the initiator is currently the owner and its neighbor was the creator.
This step prohibits malicious nodes from arbitrarily choosing which target node to communicate with, as they should first possess a legitimately acquired descriptor created by that same node in the past.
In other words, at the end of its lifetime, a descriptor serves as a certificate permitting a gossip exchange with its creator.
A correct node will never accept a gossip invitation without being presented with such a certificate, having been issued by itself in the past.

It should by now be clear that the information stored in a node descriptor is able to reveal protocol violations.
The following section explains how this information is matched to discover such illicit behavior.

\subsection{Discovering Protocol Violations}
\label{subsec:discovering-malicious-actions}

Protocol violations can be discovered by comparing two conflicting node descriptors.
To that extent, nodes should cache all descriptors they have seen in order to match them against each other and against descriptors they will receive in the future.
Note that caching a descriptor, that is, maintaining a copy of it, does not require obtaining its ownership too.
A node caching a descriptor without also owning it, is not allowed to attempt to transfer its ownership to another peer, or to attempt to redeem it.
Such an action would constitute a protocol violation.
Instead, cached descriptors are used solely for violation discovery.
We call these cached descriptors \emph{descriptor samples}, or just \emph{samples}.

When two nodes swap $s$ descriptors in a gossip exchange, they also transfer the respective ownerships to each other.
In addition to the $s$ swapped descriptors, each node sends a copy of the rest of its view to the other party, without, however, transferring the respective ownerships to it.
These descriptors will only be cached by the other node as samples.
\newline

To discover malicious actions each node should perform the following two checks for each node descriptor it receives, be it an owned descriptor or a sample:
\newline

\noindent
\textbf{Frequency Check:} The node compares the timestamp of the received descriptor against the timestamps of all stored samples created by the same node.
If the timestamps of any two descriptors of the same creator are closer to each other than the prescribed gossiping period, it constitutes indisputable proof that the creator node in question has performed a frequency violation. 

\noindent
\textbf{Ownership Check:} The node looks up its cache for a descriptor generated by the same node and with the same timestamp as the received descriptor.
If such a descriptor is found, the two samples should refer to the same original descriptor, therefore, their chains of ownership should be compatible with each other.
For instance, the two descriptors could report the same chain of ownership, e.g., $A\rightarrow B\rightarrow C$ on both.
Or, they could report different chains of ownership, with one being a subset of the other, e.g., $A\rightarrow B\rightarrow C$ on either one descriptor, and $A\rightarrow B\rightarrow C\rightarrow D\rightarrow E$ on the other.
Both these cases are perfectly ok, as they report a seemingly legitimate path the descriptor has followed during its traversing phase.

If, however, the two descriptors report ownership chains such that neither is a subset of the other, this constitutes indisputable proof of a cloning violation.
For example, if one descriptor reports $A\rightarrow B\rightarrow C\rightarrow D\rightarrow E$, and the other one reports $A\rightarrow B\rightarrow F\rightarrow G$, there is indisputable proof that $B$ illegally transferred this descriptor's ownership twice, to nodes $C$ and $F$, respectively.
\newline

If both checks pass, the node will cache the descriptor as a sample in order to compare it with descriptors received in the future.
In case two samples of the same creator and same timestamp report non-conflicting ownership chains of different lengths, the one with the longest version is retained, as it corresponds to a more updated form of that same descriptor.
If one of the checks fails, it constitutes a provable violation, and the malicious node discovered is handled as described in the following section.

\subsection{Handling Violators}
\label{subsec:disposing-malicious-nodes}

For the violators' removal problem, we take an approach similar to the one presented in~\cite{TowardsSecureEpidemics}.

Violations are discovered by finding two conflicting descriptors, both of which were previously signed by the violator.
Thus, a violation discovery is followed by the indisputable proof of the offender's identity, namely, its public key.
Presenting the two conflicting descriptors to any third node can prove to it the offender's violation and its identity.
Thus, it only takes \emph{one} node to discover a violation, for \emph{all} nodes to reliably acknowledge the fact.

When a node makes a violation discovery, it initiates a flooding to broadcast the corresponding proof to all nodes of the overlay.
Upon receiving and locally validating proofs of violation, correct nodes blacklist the corresponding malicious nodes.
Legitimate nodes immediately drop any descriptors linking them to a blacklisted node, be it descriptors they already own or ones they receive in the future.
They also stop accepting gossip requests from blacklisted nodes.
After a proof gets disseminated to the entire network, the blacklisted node effectively gets disconnected from the legitimate portion of the overlay.
After the eviction of a malicious node, any descriptors it may have cloned will gradually get replaced by fresh descriptors of other nodes.

Flooding opens a vector of DoS attacks, as malicious nodes could swarm the network with arbitrary messages, or disseminate valid proofs that refer to nodes that have already been discovered and blacklisted.
To overcome such attacks, legitimate nodes should check that each received proof has valid content, as well as that the discovered malicious node has not been yet discovered, before disseminating the proof any further.

Last but not least, proofs should be exchanged between nodes during gossip in order to inform newly joined nodes, or nodes that were absent during a flooding dissemination, about the discovered malicious nodes they are missing.


    \section{Protocol Integration}
\label{sec:protocol-integration}

The enhancements proposed by \prot\ over the original Cyclon protocol may affect certain aspects of the latter's operation.
In this section, we identify and address these cases.

\subsection{Repairing Empty View Slots}
\label{subsec:re-establishing-lost-descriptors}

In a real-world setting, losing node descriptors is inevitable.
We identify three scenarios in which descriptors may be missing from a node's view:

\begin{enumerate}
    \item When a node does not respond to a gossip request.
    This may occur due to a node failure, a network failure, or as a consequence of a non-responding malicious node.
    In this situation, the gossip initiator simply drops the unreachable node's descriptor, skips this cycle, and waits for the next cycle to initiate another gossip exchange.
    In the meantime, its view is left with one descriptor less.\

    \item In the case of an asymmetric exchange of node descriptors.
    Gossip exchanges are not guaranteed to be atomic.
    That is, it is possible for descriptor swapping to take place only in one direction, with the gossip partner failing or refusing to fulfill its part of the communication after having acquired a list of descriptors from the initiator.
    In the classic Cyclon protocol, this does not constitute a problem, as the node not receiving new descriptors is allowed to retain and reuse the descriptors it shipped to its counterparty in the incomplete exchange.
    In our approach, however, after having transferred some descriptors' ownerships to some other node, the sending node
    should discard the sent descriptors.
    Attempting to reuse the same descriptors exposes it to the risk of being accused of descriptor cloning.

    \item When a node joins the network overlay.
    Although this case does not classify as a loss of node descriptors, the state of a joining node is analogous to the state of the nodes of the aforementioned scenarios, with the distinction that a newly joined node has a completely empty view rather than a small number of empty slots in it.
    Thereby we confront the scenario of a node joining in a similar manner to the previous two scenarios.
\end{enumerate}
In all aforementioned scenarios, a node ends up in a state in which its view is not fully populated with node descriptors.

To allow nodes to populate empty slots in their views, we introduce the concept of \emph{non-swappable descriptors}.
A node with an empty slot in its view is allowed to keep a copy of a descriptor whose ownership it has transferred to some other peer, however marking it as \emph{non-swappable}.
As their name suggests, a node is not allowed to swap non-swappable links in gossip exchanges.
It can use them exclusively as gossiping tokens, that is, to redeem them to initiate an exchange with their creator when they become the oldest descriptors in its view.
Thus, a node should accept gossip invitations by nodes who present a valid non-swappable descriptor created by that node, provided the descriptor is marked as non-swappable.
The non-swappable link is redeemed, and a fresh (swappable) link of the initiator is created, thus allowing the use of non-swappable links to initiate gossiping does help in gradually repairing empty slots from nodes' views.

The non-swappable descriptor mechanism can be abused by a group of malicious nodes aiming to attack a given target node.
Specifically, such a group of nodes could pass a target descriptor's ownership around among themselves, so that they can all retain non-swappable versions of that descriptor, granting them all permission to initiate gossip exchanges with the target victim at the moment of the attack.

To tackle this malicious behavior we place the following restrictions.
First, a node must not accept more than one non-swappable link redemption for the same descriptor.
Second, a node must not accept more than one non-swappable link redemption of different descriptors in the same cycle.
Third, a node could optionally limit the number of descriptors it is willing to swap in a gossip exchange that was allowed on behalf of a non-swappable descriptor.

For newly joining nodes, it suffices to acquire some non-swappable links (e.g., through a bootstrapping procedure), in order to start gossiping and to build a healthy view, gradually populated with swappable node descriptors.

\subsection{Non-Atomic Gossip Exchanges}
\label{subsec:confronting-asymmetrical-communications}

As mentioned in the previous section, after a (malicious) node has been granted the ownership of some descriptors in a gossip exchange, it may opt not to respond, letting the other node with some empty view slots that will be filled in with non-swappable node descriptors.
This may be an explicit attack vector, which we coin the \emph{link-depletion attack}, aiming at depleting legitimate nodes of their links. 

To tackle the link-depletion attack, we alter the process of performing gossip exchanges, introducing a \emph{tit-for-tat} mechanism.
Instead of swapping all $s$ descriptors in one single message in each direction, the nodes perform $s$ round-trip communications, transferring the ownership of one descriptor at a time.
The gossip initiator goes first, by transferring its own fresh descriptor to the recipient.
If any of the nodes quits the process halfway, the other one does not send any more descriptors.
This way, the contacted node runs zero risk of losing a descriptor.
Only the initiator may end up with one descriptor too few.
However, as nodes act as gossip initiators exactly once per cycle, while they can be contacted multiple times in the same cycle, it is preferable to place the risk of losing a descriptor on the initiator rather than on the contacted node.

It should be noted that the tit-for-tat mechanism applies only to descriptor ownerships, which should be transferred one at a time.
The remaining descriptors from each node's view that are sent only as samples without transferring the respective ownerships, may be included altogether in the first message in each direction.

As presented in our evaluation, our approach in combination with the healing properties of Cyclon restricts the number of non-swappable node descriptors in the network.

\subsection{Cloning Old Descriptors}
\label{subsec:tackling-old-descriptor-duplication}

A malicious party may exploit the fact that, in Cyclon, nodes redeem the \emph{oldest} node descriptor to initiate a gossip exchange.
Specifically, when a node descriptor with a very high age is received, it is likely to get redeemed imminently, without getting the chance to be transferred to yet another node.
If that descriptor was cloned by a malicious node, its chances to be crosschecked against other clones of the same descriptor diminish.

To address this threat, we introduce a technique to our protocol referred to as the \emph{redemption cache}.
When a node redeems a descriptor, it stores a copy of that descriptor for a small number of cycles.
When gossiping, a node always sends the descriptors of its redemption cache along with the contents of its view to the other node, as sample descriptors.
The resource remands of this technique are negligible, as the redemption cache is typically configured to retain the last five or six redeemed descriptors.

    \section{Evaluation}
\label{sec:evaluation}


We performed our simulations on the \emph{PeerNet Simulator}~\cite{PeerNet}, a fork of the popular PeerSim \cite{PeerSim} simulation environment, an open-source platform for the development, testing, and deployment of P2P applications, developed in Java.

We conducted a number of simulations with 1,000 and 10,000 nodes, with view lengths ranging from 20 to 50 descriptors per node, various swap lengths, and a number of different test policies and attack scenarios.
In all experiments, there was an initialization phase, where the overlay was let emerge to a random-graph-like overlay through the protocol's self-organizing properties.

We initially perform an informal analysis of \prot's network resource requirements, and we continue with the evaluation of our protocol through simulations.

\begin{figure*}[t]
  \includegraphics{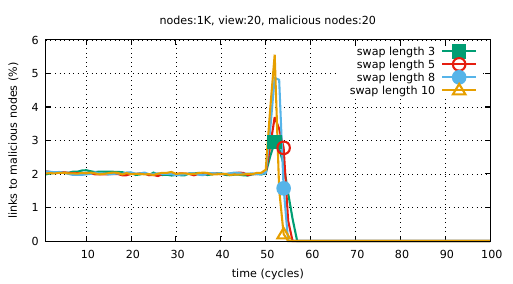}
  \includegraphics{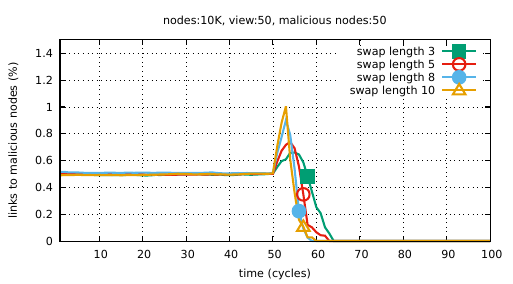} \\
  \includegraphics{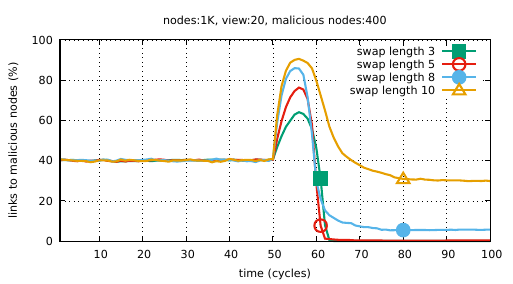}
  \includegraphics{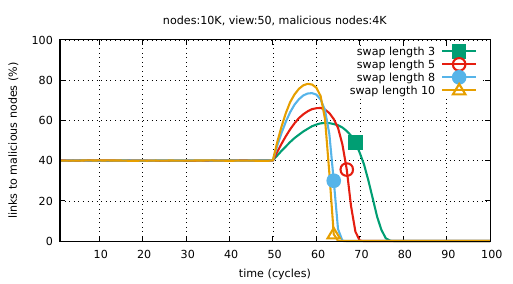}
  \caption{(Top) Same experiments as in \figref{fig:hubattack}, however this time with \prot\ shielding the overlay against adversarial behavior. (Bottom) Same experiments, however this time with malicious nodes accounting for 40\% of the network.}
  \label{fig:shielded}
\end{figure*}

\subsection{Network Costs}
\label{subsec:transmission-cost}

In our proposed configuration we choose a swap length $s=3$, as it is sufficient for fast convergence in Cyclon, and a view length $\ell=20$.
We also set the redemption cache (see \secref{subsec:tackling-old-descriptor-duplication}) to hold $r=5$ descriptors.

The node information contained in a descriptor, namely its public key (256 bits), IP address (32 bits), port (16 bits), and timestamp (64 bits), accounts for a total of 368 bits.
Each time the descriptor is transferred over to a new owner, an extra public key (256 bits), and a signature (256 bits) are appended to it.
Therefore, a descriptor's size is $368 + 512 \cdot t$ bits, where $t$ is the number of times its ownership has been transferred.

As stated in \cite{voulgaris.jnsm.2005}, each descriptor lives for an average of $\ell$ cycles before getting redeemed, where $\ell$ is the view length.
As a node participates on average in two gossip exchanges per cycle (one it initiates and one it is invited to), and in each gossip exchange it transfers $s$ out of $\ell$ descriptors to its gossip partner, a descriptor's chance to be transferred is $\frac{s}{\ell}$ per gossip, thus $\frac{2s}{\ell}$ per cycle.
Therefore, in its lifetime of $\ell$ cycles a descriptor will have been transferred $2s$ times, on average.

Taking a pessimistic scenario of all descriptors having been transferred $2s=6$ times (pessimistic, as younger descriptors have not been transferred as much yet), we get a back-of-the-envelop estimate of a descriptor size being $368+512*6 = 3440$ bits, or 430 Bytes.
As in each gossip exchange each peer sends to its counterparty $\ell+r = 25$ descriptors, we conclude that a gossip exchange involves the transfer of roughly 10.5 KBytes in each direction.

This is a negligible network overhead, especially when considering that typical gossiping periods would be somewhere between 10 and 60 seconds.

\begin{figure*}[t]
  \begin{tabular}{cc}
    \includegraphics{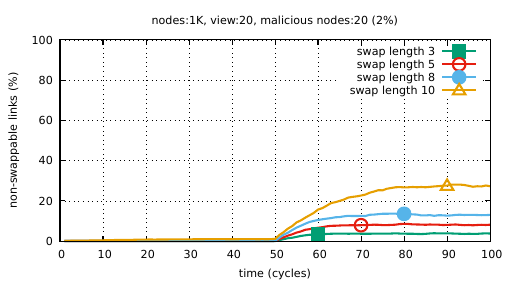} &
    \includegraphics{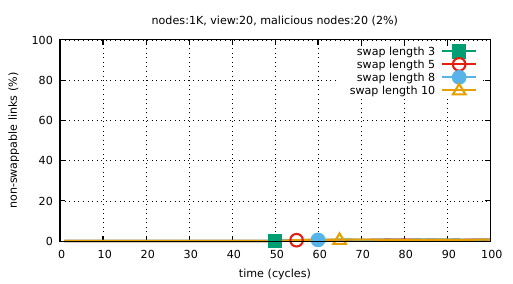}  \\
    \small{Tit-for-tat: disabled} & \small{Tit-for-tat: enabled} \\
    \includegraphics{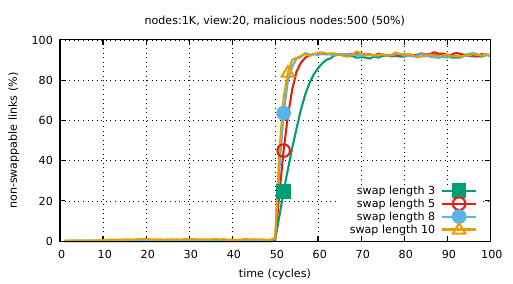} &
    \includegraphics{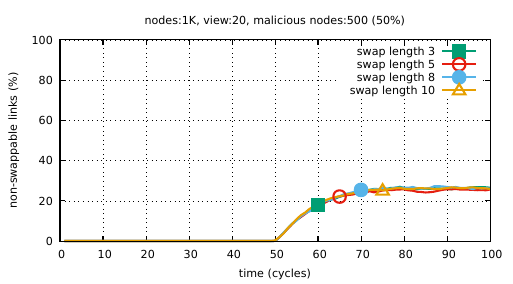}  \\
    \small{Tit-for-tat: disabled} & \small{Tit-for-tat: enabled}
  \end{tabular}
  \caption{The link-depletion attack: Malicious nodes present an empty view to legitimate ones, in order to deplete their views of their descriptors. Its effect on the percentage of non-swappable descriptors when tit-for-tat is disabled (left) or enabled (right).
  The effectiveness of the tit-for-tat mechanism is proven for a limited number of malicious nodes (top), and for the adverse scenario where $50\%$ of the nodes are malicious (bottom). The attack starts after cycle 50, at a converged overlay state.}
  \label{fig:swappable}
\end{figure*}

\subsection{Defending Against The Hub Attack}
\label{subsec:defending-against-hub}

To assess the effectiveness of \prot\ against a hub attack, we assume a modified version of the attack described in \secref{sec:challenges}.
In this attack, malicious nodes maintain a central pool of descriptors, comprising copies of all the descriptors generated by malicious nodes in recent cycles.
When gossiping with a legitimate peer, a malicious node presents a fake view consisting exclusively of descriptors to other malicious nodes, selected out of this central pool.


We also assume that all legitimate nodes participate in the \prot\ protocol.
When a legitimate node discovers a protocol violation, it broadcasts the proof over the links it has acquired through \prot.

We first examine the scenario where the smallest group of malicious nodes that could hijack an overlay attempts an attack.
Such a group should have a minimum of $\ell$ malicious nodes, as having just one node less would leave space for legitimate nodes to also maintain legitimate links.

\figref{fig:shielded}(top) presents two such attempts: one for a network of 1K nodes, $\ell=20$, and 20 malicious nodes ($2\%$ of the nodes), and one for 10K nodes with $\ell=50$ and 50 malicious nodes ($0.5\%$ of the nodes).
Malicious nodes start the attack on cycle 50.
Up until that cycle, the percentage of legitimate nodes' links to malicious nodes is proportional to their population (i.e., $2\%$ and $0.5\%$, respectively), as expected.
After cycle 50, a small spike begins to form, as malicious nodes have aggressively started to pollute the overlay.
This trend stops abruptly and the percentage of legitimate nodes' links to malicious nodes decreases rapidly, as malicious nodes get detected and removed from the overlay.

This figure should be compared to \figref{fig:hubattack}, which corresponds to the same attack on legacy Cyclon, without \prot's security measures in place.

\figref{fig:shielded}(bottom-left) explores how our protocol performs in an extreme attack scenario, where the adversary controls $40\%$ of the node population.
As shown, malicious link population is temporarily increased, reaching values from $60\%$ up to $90\%$ of the total links, depending on the swap-length parameter.
A few cycles later, however, it decreases rapidly, as malicious nodes get purged from the network.
It can be noticed that in the cases where the swap length is very high in comparison to the view length, malicious links may not get completely eliminated.
For $s=8$ and $s=10$, the ratio of malicious links does not drop below $7\%$ and $30\%$, respectively.
This suggests that the respective percentages of legitimate nodes have been eclipsed due to their exposure to a high number of malicious descriptors.
Specifically, $7\%$ and $30\%$ of the legitimate nodes have their links controlled by malicious nodes, rendering them unable to receive any new malicious discovery proofs through dissemination.

In \figref{fig:shielded}(bottom-right), it can be observed that despite employing the same swap length as in the previous experiment, no nodes get eclipsed.
This is due to the fact that the swap length is now far smaller than the view length, making it more challenging for the view to be filled with malicious descriptors.
Thus, it is important to pick a sufficiently low swap length in order to be able to withstand malicious attacks that involve a high number of malicious nodes.
Generally, as shown in~\cite{voulgaris.jnsm.2005}, a swap length of 3 is sufficient for fast convergence even in networks with a high number of nodes.

Concluding, we performed the aforementioned experiments with $50\%$ of the nodes being malicious, and we observed that the legitimate nodes manage to effectively tackle the attack if a swap length of 3 is adopted.

\subsection{Tit-for-Tat Evaluation}
\label{subsec:integration-evaluation}

To evaluate the tit-for-tat communication mechanism proposed in \secref{subsec:confronting-asymmetrical-communications}, we expose our nodes to the most effective attack that could happen in this regard.
During a gossip exchange, a malicious node is transmitting an empty view in response to the list of descriptors provided by the legitimate node. 
The goal of such an attack is to deplete legitimate nodes of their links, effectively letting them with non-swappable links, and rendering the network static.

The attack is depicted in \figref{fig:swappable}, for a network size of 1,000 nodes, with a view length of 20, and various swap lengths.
Malicious nodes start deploying this attack at cycle 50.
The figures on the left show the behavior without the tit-for-tat mechanism, while the figures on the right incorporate tit-for-tat.

We first examine a scenario where malicious nodes account for $2\%$ of all nodes, depicted in \figref{fig:swappable}(top).
After the attack starts, we see a number of non-swappable links proportional to the swap length forming in the left plot, which is reasonable, as with a higher swap length a legitimate node can lose more node descriptors per gossip exchange.
The figure on the top right side clearly demonstrates the efficiency of the tit-for-tat mechanism, that effectively minimizes link depletion to a negligible level.

As a second set of experiments, we study the scenario where malicious nodes constitute half of the total node population, depicted in \figref{fig:swappable}(bottom).
Under the observed scenario, irrespectively of the swap length, nearly all the links in the views of legitimate nodes become non-swappable.
The healing mechanism of Cyclon enables the retention of a muscle fraction of links as swappable, approximately $4\%$ of the total.
Even in such a detrimental situation, as depicted in the bottom right plot, the tit-for-tat mechanism successfully limits the proportion of non-swappable links to approximately $27\%$.



\subsection{Redemption Cache Evaluation}

\begin{figure*}[t]
  \includegraphics{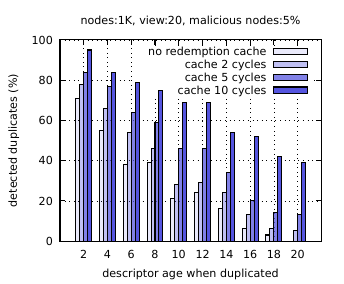}
  \includegraphics{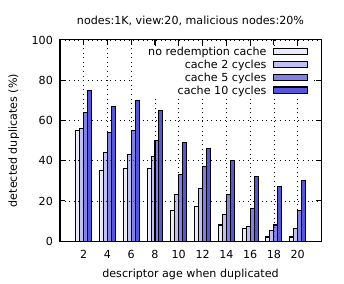}
  \includegraphics{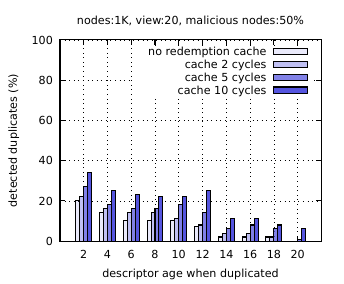}
  \caption{
    Detection ratio as a function of (i) the link's age at duplication time, (ii) the redemption cache size, and (iii) the percentage of malicious nodes in the network.
  }
  \label{fig:detections}
\end{figure*}

Concluding, we evaluate our redemption cache mechanism, described in \secref{subsec:tackling-old-descriptor-duplication}.
\figref{fig:detections} illustrates the correlation between the age at which a descriptor was cloned and its respective detection ratio, for various percentages of malicious node participation and redemption cache sizes.

We observe that, for descriptors duplicated at a low age, the transmission of views is sufficient to detect them with high probability.
For descriptors duplicated at a higher age, the respective detection ratio drops, and the redemption cache's role becomes evident.

For the scenario of 5\% and 20\% malicious participation, we observe that a redemption cache of just $r=5$ descriptors is able to detect more than 20\% of the illegal duplications.
Things get worse when the malicious participation takes extreme values, as showcased in the third figure, including a malicious population of 50\% of the network.
When considering such cases, the redemption cache should be set to a size of 10 descriptors, in order to be able to detect approximately 5\% of duplications.

It should be emphasized that a malicious node should only be found guilty once in order to be deterministically evicted from the network, thus even with a detection ratio of 5\%, malicious nodes will eventually be detected and dropped out.


    \section{Related Work}
\label{sec:related-work}

In~\cite{TowardsSecureEpidemics} a general methodology for discovering illicit actions and identifying malicious nodes
in gossip protocols is described.
The solution proposed utilizes signed communication records in order to hold nodes accountable for their actions.
The consistency of a node's communication history is verified through anonymous auditioning.
Apart from the discovery of frequency violations, it is not clear how this solution could be adjusted in order to resolve
violations in peer sampling protocols.
Furthermore~\cite{TowardsSecureEpidemics} relies heavily on anonymous auditions of nodes in addition to the standard protocol
communication, which may prove unreliable in certain cases, 
and which increases the overall bandwidth costs.

A series of works have particularly attempted to tackle the Hub Attack~ \cite{SecurePeerSamplingService}, \cite{IdentifyingMaliciousPeers}, \cite{SecurePeerSampling}, \cite{PuppetCast}.
In~\cite{SecurePeerSamplingService} and~\cite{IdentifyingMaliciousPeers}, each node attempts to profile the nodes it interacts with by comparing the descriptors they receive with those present in their own views.
The underlying rationale behind this approach is that since malicious behavior often revolves around duplicating peers, if a received set of descriptors shares a significant number of common descriptors with a node's view, it strongly implies that the sender is malicious with high probability.
In Secure Peer Sampling~\cite{SecurePeerSampling}, nodes attempt to estimate the indegrees of their neighbors by performing extra communication steps.
When a node is estimated to have an indegree that is considerably higher than the estimated average indegree of all nodes, it undergoes local blacklisting.

In the aforementioned approaches, it is not possible to deterministically establish the malicious nature of nodes.
Additionally, due to the possibility of false-positives, a malicious node is never completely held accountable for its actions.
On the contrary, \prot\ strives to uncover and remove malicious nodes, unburdening the overlay from further malicious behavior and impactful attack attempts.
Furthermore, because in the aforementioned approaches malicious nodes are continuously presented with a second chance, the pollution in the overlay never declines to zero.

In PuppetCast \cite{PuppetCast}, each node possesses a static view that is issued by a trusted authority, and a mutable view that is continuously altered by inserting descriptors from other nodes' static views.
This approach effectively addresses the view violations described in Section~\ref{sec:challenges}, as nodes exchange only the static views that are certified by the central authority.
On the other hand, the adoption of central nodes opens a new vector of attacks and could introduce performance bottlenecks.
Our approach, on the contrary, is completely decentralized.

Brahms~\cite{Brahms} adopts a different strategy in which every node establishes an unbiased random set of neighbors by applying a number of independent permutations on node IDs that have been previously observed by the node.
Consequently, only a small part of the view of each node comprises unbiased descriptors, with the rest of it
consisting of descriptors collected through gossiping without any guarantees.

As demonstrated in Raptee~\cite{RAPTEE}, despite the fact that~\cite{Brahms} prevents partitioning, it fails to restrict the presence
of duplicated node descriptors.
Raptee achieves a drastic reduction in the prevalence of malicious over-representation observed in~\cite{Brahms}, by utilizing nodes with attested computing capabilities provided by a specific software guard extension.
It does so by allowing attested nodes to generate a higher quantity of node descriptors compared to non-trusted nodes.
Our approach exceeds Raptee as it eliminates the number of malicious node descriptors instead of bounding them.

By categorizing peers based on their IP prefixes, HAPS~\cite{HAPS} instructs nodes to pick neighbors with a high diversity of IP prefixes.
The underlying rationale behind~\cite{HAPS} is based on the notion that the acquisition of IP addresses with diverse IP prefixes entails significant costs, rendering it difficult for a malicious party to consolidate a large number of descriptors in a single node's view.

    \section{Conclusions}
\label{sec:conclusions}

We presented \prot, a protocol that extends the legacy Cyclon peer-sampling protocol to shield it against arbitrary protocol violations.
\prot\ leverages the continuous exchange of node descriptors that is inherent in Cyclon, in order to disseminate evidence of malicious actions.
Such evidence can be utilized by correct nodes to construct indisputable proof of illicit actions, and to inadvertently expose the identity of the respective offenders, allowing legitimate nodes to blacklist them, protecting the network from their malicious actions.

Our evaluation demonstrates that \prot\ does not only prevent attacks with a reasonable malicious node participation, but also copes well with massive-scale attacks, as it gradually exposes and eliminates malicious nodes, blocking them from the legitimate part of the network.

To the best of our knowledge, this is the first gossip-based peer-sampling protocol that not only confronts the problem of arbitrary protocol violations, but also exposes perpetrators with provable evidence of their malicious actions.

    \section*{Acknowledgments}

This work was funded by Input Output (\url{iohk.io}) in the context of the \emph{``Eclipse-Resistant Network Overlays for
Fast Data Dissemination''} project, aiming at optimizing and securing Cardano's network overlay.
We gratefully thank Input Output for their support, and their engineering team for fruitful discussions and valuable feedback.

    \bibliographystyle{IEEEtran}
    \bibliography{./paper}

\end{document}